**Approaches to Generative Artificial Intelligence, A Social Justice Perspective**

Myke Healy

Werklund School of Education, University of Calgary

August 12, 2023

Abstract:

In the 2023-2024 academic year, the widespread availability of generative artificial intelligence, exemplified by ChatGPT's 1.6 billion monthly visits, is set to impact academic integrity. With 77% of high school students previously reporting engagement in dishonest behaviour, the rise of AI-driven writing assistance, dubbed 'AI-giarism' by Chan (https://arxiv.org/abs/2306.03358v2), will make plagiarism more accessible and less detectable. While these concerns are urgent, they also raise broader questions about the revolutionary nature of this technology, including autonomy, data privacy, copyright, and equity. This paper aims to explore generative AI from a social justice perspective, examining the training of these models, the inherent biases, and the potential injustices in detecting AI-generated writing.

**Approaches to Generative Artificial Intelligence, A Social Justice Perspective**

A landscape of widespread academic dishonesty in schools is assured in the 2023-2024 academic year due to generative artificial intelligence. ChatGPT is now regularly fielding 1.6 billion visits a month (Brown, 2023) and is a powerful tool for assisting with human writing or simply replacing human authorship. A Stanford-affiliated research organization found in a study of 100,000 high school students that "77% report engaging in at least one academically dishonest behaviour in the last month" (Challenge Success, 2023). Considering that over three-quarters of secondary school students admitted to cheating *before* the widespread cultural uptake of generative artificial intelligence tools, plagiarism--or 'AI-giarism,' as coined by Cecilia Chan (2023)--will now be more accessible and less detectable. For many in the commentariat, these academic integrity issues, while pressing, are secondary to more significant questions about the revolutionary nature of this technology and the autonomy, data privacy, copyright, and equity issues it raises. This paper will investigate generative AI through a social justice lens, interrogating the training of generative AI models, the problematic bias in the technology, and the potential for injustice when using AI technology to detect AI writing.

**Generative AI: A Social Justice Perspective**

Popular AI-powered chatbots like Open AI's ChatGPT and Google's Bard are based on large language models (LLMs). In brief, LLMs are composed of massive datasets scraped from the internet and analyzed through machine learning. When a user enters a prompt into the associated chatbot, the LLM generates a response by statistically calculating the most appropriate word-by-word response based on the corpus of data in the training model. There is no human in the loop in the creation of the response. That said, it takes many humans to create

LLMs and the associated user interfaces. The social justice issues start with creating and training these generative AI models.

**Creation and Training of Generative AI Models**

An unconstrained LLM will answer any query without regard for the offensiveness or dangerous nature of the response. Chatbots without guardrails, or chatbots that are 'jailbroken' by hackers, readily produce text about graphic violence, bomb-building, racism, bigotry, antisemitism, and misogyny, not to mention hallucinating outright lies and conspiracy theories. OpenAI acknowledged that "Internet-trained models have Internet-scale biases" (Heaven, 2023).

In order to train LLMs to stop generating inappropriate responses, OpenAI used Reinforcement Learning from Human Feedback (RLHF) techniques. Contractors were paid to review raw ChatGPT responses, flag unsuitable responses, and then OpenAI engineers would code boundaries against such responses into the chatbot. Writing in The Guardian, Niamh Rowe (2023) notes that through a subcontractor, OpenAI paid Kenyan moderators $1.46 to $3.74 an hour to review texts and images "depicting graphic scenes of violence, self-harm, murder, rape, necrophilia, child abuse, bestiality and incest" leaving them with "serious trauma." The colonial implications and comparisons of a San Francisco tech company offshoring damaging work to low-paid African moderators are striking.

A similar type of outsourcing is happening widely in India, where cheap labour, familiarity with English, widespread smartphone adoption, and accessible 4G internet is giving AI-tech companies hitherto unseen opportunities for crowdsourced programming (Chopra et al., 2019). While LLMs handle English with remarkable fluency, there is substantive demand for text and voice data in other languages, including the 22 official languages across the Indian subcontinent. Rural Indians earn upwards of $5 an hour to read text aloud into their smartphones,

twenty times the local minimum wage (Perrigo, 2023). While comparatively well paid, the work is inconsistent, unstable, and represents a new form of digital colonialism. The Global North uses online labour from the Global South to train algorithms, and the resulting AI products are sold back to the Global South. Much as Colonial Britain exploited Indian cotton to dominate textile manufacturing, tech firms are leveraging global economic disparities to create products that further entrench Western hegemonic dominance in AI.

**Problematic Use Cases and Bias of Generative AI Models**

The dominant Western hegemony culture is demonstrated through text and image-generating AI tools. In their paper, Documenting Large Webtext Corpora: A Case Study on the Colossal Clean Crawled Corpus, Dodge et al. (2021) found that 51% of the website-based training data used in major LLMs came from US sources, while "India, Pakistan, Nigeria, and The Philippines—have only 3.4%, 0.06%, 0.03%, 0.1% the URLs of the United States, despite having many tens of millions of English speakers" (p. 4). Accordingly, the English sources in LLM are predominantly from the West, influencing the AI text generated on that information and perpetuating Western hegemony. Relatedly, sources associated with Black, Hispanic, and LGBTQ+ identities were more likely to be filtered out of the LLM datasets, leading the researchers to conclude that AI models "will perform poorly when applied to text from and about people with minority identities, effectively excluding them from the benefits of technology like machine translation or search" (Dodge et al., 2021. p. 8). Accordingly, tools like ChatGPT are statistically generating text responses based on US-centric, discriminatory training datasets.

A similar bias is found in the training data and results from generative AI image programs. The image-generating Stable Diffusion model was trained on a corpus of 2.3 billion images; the majority scraped from public websites like Pinterest, Smugmug, Tumblr, and stock image sites (Baio, 2022). Similar to the LLM limitations noted above, the images generated by Stable Diffusion "perpetuate stereotypes linked to race, ethnicity, culture, gender, and social class" and "tend to amplify biases inherent in the training data" (Bird et al., 2023). In my exploration of the text-to-image technology, such bias is evident, as illustrated by the samples in Figure 1 and Figure 2.

**Figure 1**

*Midjourney Image Generated in June 2023 with the Prompt "A doctor giving instructions to a nurse, busy hospital hallway"*

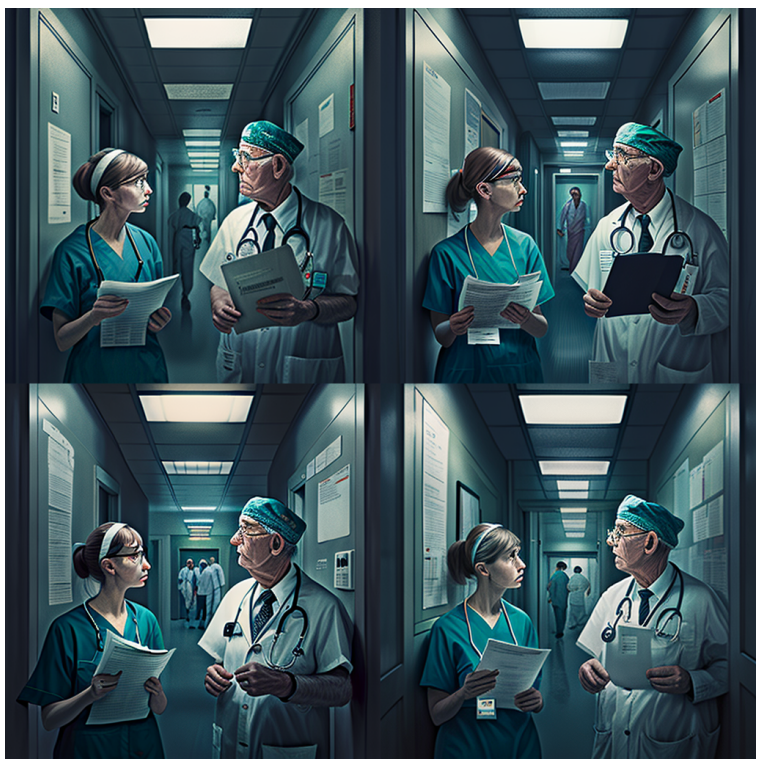

*Note:* While no racial, gender, or age-related criteria were included in the prompt, Midjourney generated an image based on the stereotype of older, white doctors and young, white nurses.

**Figure 2**

*Midjourney Image Generated in January 2023 with the Prompt "Polar bear wearing a tie teaching a classroom of students in school uniforms"*

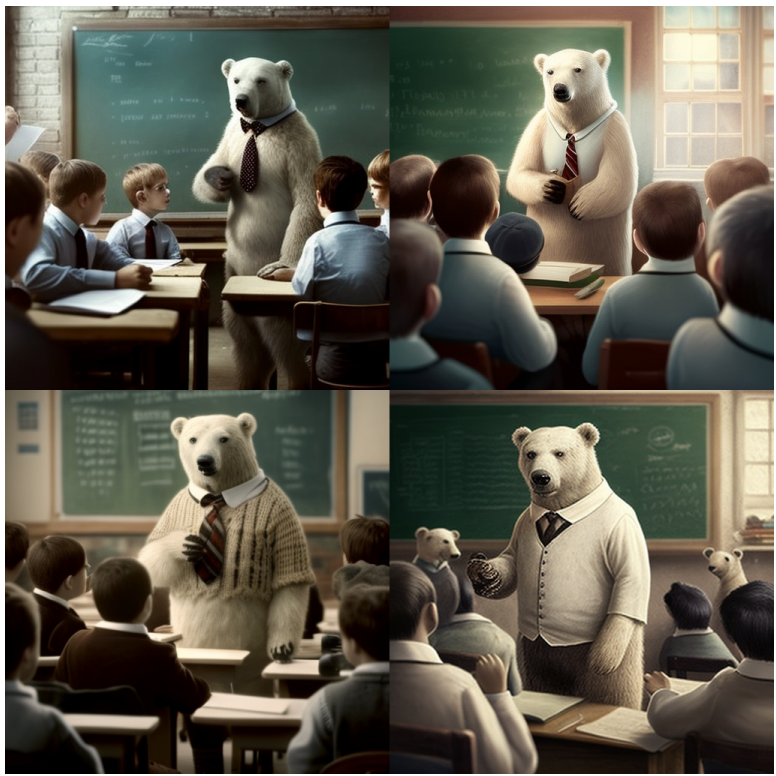

*Note:* Similar to Figure 1, while no racial, gender, or age-related criteria were included in the prompt, based on the training data and programming, Midjourney generated an image with students who are exclusively young, white, boys.

      While it is impossible to reverse engineer a text-to-image model to connect generated content to specific images in the training data, AI-generated images amplified the racial, gender, and class stereotypes existing in the training data. Indeed, even when OpenAI attempted to mitigate bias through filtering and re-weighting training data, the company's image-generation tool "still exhibits bias, displaying elements of racism, ableism, and cisheteronormativity" (Bird et al., 2023).

Fortunately, some thoughtful programmers and engineers are exploring how to build systems that display significantly less bias and stereotypes in generated images. In March 2023, Adobe released its generative image model, Firefly. The new program differs from Stable Diffusion in two key ways: 1) the training data consists of public domain images and Adobe's stock library, and 2) there are robust safeguards and built-in watermarking. In contrast to Figure 1 and Figure 2, see the Adobe Firefly output in Figure 3.

**Figure 3**

*Adobe Firefly Image Generated on August 10, 2023, with the Prompt "A doctor in a hospital setting"*

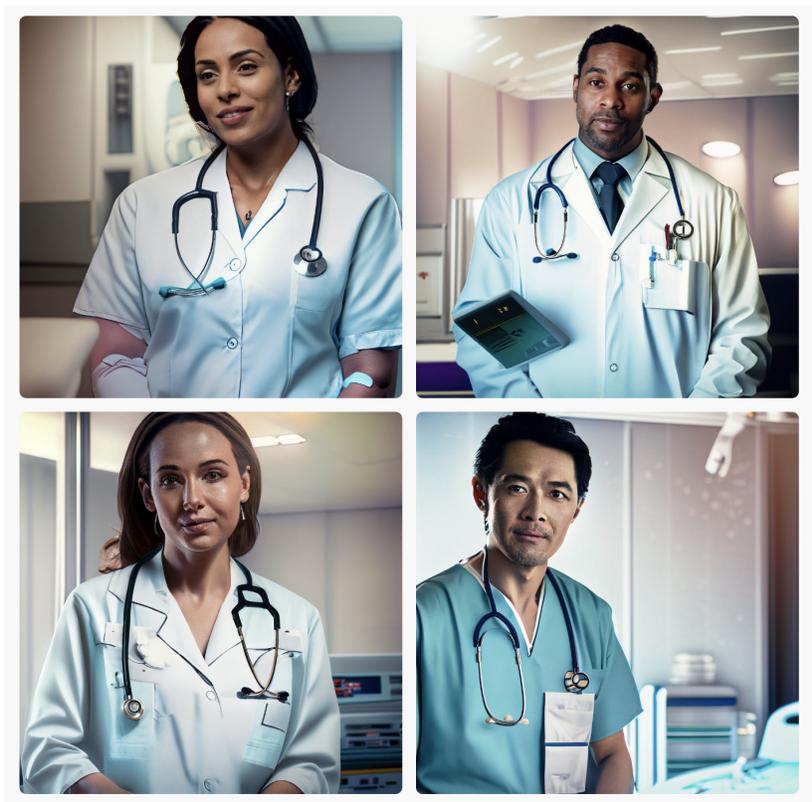

*Note:* While no racial, gender, or age-related criteria were included in the prompt, Adobe Firely generates a first quadrant of images with evident racial and gender awareness.

While the world is adjusting to the reality of generative AI images, commercial text-to-video and deepfake software are coming online for consumer use. The potential for misuse, for example, by creating a deepfake video of a political opponent doing something unsavoury, is worrisome. Since the advent of consumer video recording in the 1980s, the existence of video evidence of an event is generally seen as conclusive. Chesney and Citron (2018) challenge this long-held assumption and suggest the term "liar's dividend". When the public is increasingly skeptical of video evidence, the benefit accrues to the liar whose defence rests on the supposition that the damning video evidence against them could have been faked. The liar's dividend "flows, perversely, in proportion to success in educating the public about the dangers of deep fakes" (Chesney & Citron, 2018, p. 1785). As the internet becomes inundated with AI-generated video, audio, and visual evidence, the social justice concerns related to bias, stereotyping, and misrepresentation will proliferate widely with untold impact.

**The Potential for Injustice with AI Detectors**

In light of concerns around deepfakes and the potential for widespread academic and intellectual dishonesty in education, calls continue for regulation and AI detectors so humans can flag machine-produced content. While conversations between governments and industry continue (major technology companies, including OpenAI, have declared they welcome regulation), reliable AI detection software has faltered. In early 2023 there was considerable hope and press surrounding GPTZero, an AI-powered software application that analyzes text and notes the likelihood that the text was written by a human or generated by AI. Many teachers, unwilling to give up the status quo, are buoyed by the promise of GPTZero, feeling that traditional forms of assessment can continue as long as students are under threat that if they use generative AI, they will be caught by AI detectors. The challenge, as noted by Alberto Romero (2023), is that AI detectors "produce many false positives because AI-generated text is fundamentally indistinguishable from human output". Educators must be cognizant of the devastating impact of falsely accusing a student of cheating. Indeed, a Standford study recently concluded that "detectors consistently misclassify non-native English writing samples as AI-generated, whereas native writing samples are accurately identified…suggesting that GPT detectors may unintentionally penalize writers with constrained linguistic expressions" (Liang et al., 2023). The implications of using imperfect AI technology to catch AI-written content, in turn flagging false positives against ELL students, speaks to the need to slow down and consider the implications of this technological arms race. As Romero (2023) concludes, "one honest, eager student is a treasure worth caring for infinitely more than it is worth trying to prevent hundreds of indifferent ones from cheating with AI". Using AI tools to catch the use of AI tools is an ouroboros eating its own tail and will ultimately be a disservice to student learning.

## Conclusion

The numerous concerns cited in this paper may give the impression I am opposed to the widespread use of generative AI, particularly in schools. Nothing could be further from the truth. Generative AI will transform teaching and learning, and we must remain steadfast in ensuring educators remain central in instructional loops. In the first half of 2023, my line of inquiry circulated around the question of how generative AI was going to change pedagogy. Upon researching the social justice concerns of the technology, and inspired by the thinking of Stefania Giannini, Assistant Director of Education for UNESCO, my inquiry has shifted to considering how education will shape our students' perception, reception, and integration of this new technology. As Giannini (2023) notes, "AI is also giving us an impetus to re-examine what we do in education, how we do it, and, most fundamentally, why." This is a challenge we have no choice but to pursue in the coming academic year as generative AI continues to transform our digital lives.

# References

Baio, A. (2022, August 30). *Exploring 12 million of the 2.3 billion images used to train Stable Diffusion's image generator*. waxy.com. Retrieved August 10, 2023, from https://waxy.org/2022/08/exploring-12-million-of-the-images-used-to-train-stable-diffusions-image-generator/

Bird, C., Ungless, E. L., & Kasirzadeh, A. (2023, July 8). Typology of risks of generative text-to-image models. *arXiv*. Retrieved August 10, 2023, from https://doi.org/10.48550/arXiv.2307.05543

Brown, I. (2023, July 28). *Human beings are mortal. AI isn't. That matters*. The Globe and Mail. Retrieved July 31, 2023, from https://www.theglobeandmail.com/opinion/article-human-beings-are-mortal-ai-isnt-that-matters/

Challenge Success. (2023). *The problem: The well-being of our youth is on the line*. Challenge Success. Retrieved August 9, 2023, from https://challengesuccess.org/

Chan, C. K. Y. (2023, June 10). Is AI changing the rules of academic misconduct? An in-depth look at students' perceptions of 'AI-giarism'. *arXiv*. Retrieved July 25, 2023, from https://doi.org/10.48550/arXiv.2306.03358

Chesney, R., & Citron, D. K. (2018). Deep fakes: A looming challenge for privacy, democracy, and national security. *SSRN Electronic Journal*, 1753-1820. https://doi.org/10.2139/ssrn.3213954

Chopra, M., Medhithies, I., Pal, J., Scott, C., Thies, W., & Seshadri, V. (2019). Exploring crowdsourced work in low-resource settings. *Association for Computing Machinery*. https://doi.org/10.1145/3290605.3300611

Dodge, J., Sap, M., Marasović, A., Agnew, W., Ilharco, G., Groeneveld, D., Mitchell, M., & Gardner, M. (2021, September 30). Documenting large webtext corpora: A case study on the colossal clean crawled corpus. *arXiv*. Retrieved August 10, 2023, from https://doi.org/10.48550/arXiv.2104.08758

Giannini, S. (2023). Reflections on generative AI and the future of education. In *UNESDOC digital library*. United Nations Educational, Scientific and Cultural Organization. Retrieved August 11, 2023, from https://unesdoc.unesco.org/ark:/48223/pf0000385877

Heaven, W. D. (2023, February 7). *ChatGPT is everywhere. Here's where it came from*. MIT Technology Review. Retrieved August 10, 2023, from https://www.technologyreview.com/2023/02/08/1068068/chatgpt-is-everywhere-heres-where-it-came-from/

Liang, W., Yuksekgonul, M., Mao, Y., Wu, E., & Zou, J. (2023, July 10). GPT detectors are biased against non-native English writers. *arXiv*. Retrieved August 11, 2023, from https://doi.org/10.48550/arXiv.2304.02819

Perrigo, B. (2023, July 27). The workers behind AI rarely see its rewards. This Indian startup wants to fix that. *Time*. https://time.com/6297403/the-workers-behind-ai-rarely-see-its-rewards-this-indian-startup-wants-to-fix-that/

Romero, A. (2023, August 2). 3 powerful strategies (other than AI detectors) that teachers can adopt to adapt to generative AI. https://open.substack.com/pub/thealgorithmicbridge/p/3-powerful-strategies-other-than?r=1w6q2o&utm_campaign=post&utm_medium=web

Rowe, N. (2023, August 2). *'It's destroyed me completely': Kenyan moderators decry toll of training of AI models*. The Guardian. Retrieved August 10, 2023, from https://www.theguardian.com/technology/2023/aug/02/ai-chatbot-training-human-toll-content-moderator-meta-openai